\documentclass[%
,secnumarabic%
,tightenlines%
,amssymb, amsmath,nobibnotes, aps,nofootinbib,showpacs,]{revtex4}
\usepackage{epsfig}%
\usepackage{graphicx}%

\begin{document}

\title{Thermal relics in cosmology with bulk viscosity}%

\author{A. Iorio$^a$}

\author{G. Lambiase$^{b,c}$}

\affiliation{$^a$Faculty of Mathematics and Physics, Charles University in Prague, V Holesovickach 2,
18000 Prague 8, Czech Republic}
\affiliation{$^b$University of Salerno, 84084 - Fisciano (SA), Italy.\\
$^c$INFN, Sezione di Napoli, Italy,}
\date{\today}%
\def\be{\begin{equation}}
\def\ee{\end{equation}}
\def\al{\alpha}
\def\bea{\begin{eqnarray}}
\def\eea{\end{eqnarray}}

\renewcommand{\theequation}{\thesection.\arabic{equation}}

\begin{abstract}

In this paper we discuss some consequences of
cosmological models in which the primordial cosmic matter is described by a relativistic imperfect fluid.
The latter takes into account the dissipative effects (bulk viscosity) arising from different cooling rates
of the fluid components in the expanding Universe. We discuss, in particular, the effects of the bulk viscosity on
Big Bang Nucleosynthesis and on the thermal relic abundance of particles, looking at recent results of PAMELA
experiment. The latter has determined an anomalous excess of positron events,
that cannot be explained by the conventional cosmology and particle physics

\end{abstract}

\maketitle

\section{Introduction}
\setcounter{equation}{0}

Bulk viscous pressure in cosmic media arises as a consequence of the coupling among the different components of the cosmic substratum.
Since these components have different internal equation of state, their cooling is different as the Universe expands, giving rise
to a deviation of the system from equilibrium. The different cooling rates of the components generate a bulk viscous pressure of the cosmic medium as a whole, that
for homogeneous and isotropic Universe is the only possible dissipative phenomenon\footnote{More specifically, bulk viscosity effects account for the rapid expansion or compression of a fluid ceasing to be in thermal equilibrium. As a consequence, the bulk viscosity gives a measure of the pressure
that is necessary for restoring the equilibrium to a compressed or expanding system. This condition arises in a natural way in a cosmological
expanding background, such as the early Universe.}. These dissipative processes are described in terms of imperfect fluid (see \cite{Zimdahl2001,Zimdahl2007,Zimdahl1998,Zimdahl1998a,Zimdahl1996,Zimdahl1999,Sawyer,Wolvaardt,barrow,fuller,stoker,brevik,haslam,
zerbini,setare,harko,odintsov,velten,singh,devecchi,elizalde,pavon,fabris,marteensreview} and references therein).
In a FRW Universe, the dissipation is a scalar and can be therefore described as a bulk viscosity as referred to the thermodynamical approach
\cite{marteensreview}.

Since bulk viscosity enters into the Einstein field equations, it is natural to expect that it may affects the evolution of the primordial Universe. Such a {\it modification} of the standard cosmology alters the thermal histories of (relic) particles, hence their abundance. This scenario can be realized {\it before} the Big Bang Nucleosynthesis (BBN) epoch, a period of the Universe evolution not directly constrained by cosmological observations. When the expansion rate of the Universe is enhanced as compared to that one derived in the framework of general relativity, thermal relics decouple with larger relic abundance. The change in the Hubble rate may have therefore its imprint on the relic abundance of dark matter, such as WIMPs, axions, heavy neutrinos, and so on.
These studies are strongly motivated by recent astrophysical results which involve cosmic ray electrons and positrons \cite{PAMELA6,ATIC7,Fermi-LAT8,HESS9-10}, antiprotons \cite{PAMELA11}, and $\gamma$-rays \cite{HESS12,Fermi-LAT13-14}. Particular attention is devoted to the presence of the peak in the cosmic
positron spectrum at energies above 100 GeV observed in PAMELA experiment \cite{PAMELA6} (see also \cite{arkani,robertson,cirelli,blasi}).
Moreover, in this paper the effects of the viscous Universe are also investigated in the framework of the Big Bang Nucleosynthesis (BBN), i.e. the primordial light element abundance such as ${}^3 He$, ${}^4 He$, $D$, ${}^7 Li$, and in the framework of matter-antimatter asymmetry in the Universe referring in particular to the so called gravitational baryogenesis, a mechanism that gives rise to the origin of matter-antimatter asymmetry through a coupling between the baryon/lepton currents and the scalar curvature of the Universe.

The paper is organized as follows. In Section 2 we recall the main features of the bulk viscosity. In Section 3 we derive the constraints provided by Big Bang Nucleosynthesis. Section 4 is devoted to the analysis of relic thermal abundance, and we derive the order of magnitude of the mass of the dark matter particles required to explain the PAMELA experiment. In Sections 5 we discuss an interesting aspect of bulk viscosity in connection with the origin of matter-antimatter asymmetry in the early Universe. Conclusions are discussed in Section 6.

\section{Imperfect fluid}
\setcounter{equation}{0}

The general form of the energy-momentum tensor is \cite{zimdahl2000}
 \be\label{energy-momentum-tensor}
 T_{\alpha\beta}=\rho u_\alpha u_\beta +(p+\Pi)h_{\alpha\beta}+q_\alpha u_\beta+q_\beta u_\alpha +\pi_{\alpha\beta}\,,
 \ee
where $\Pi$ is the scalar pressure (the bulk viscous pressure), $h_{\alpha\beta}=g_{\alpha\beta}+u_\alpha u_\beta$ is the projector tensor,
$q^\alpha$ is related to the energy flux, $\pi_{\alpha\beta}$ is the anisotropic stress. $q^\alpha$ and $\pi_{\alpha\beta}$ satisfy the relations
 \[
 q^\alpha u_\alpha =0\,, \quad \pi_{\alpha\beta}u^\beta = 0 = \pi^\alpha_{\,\,\,\alpha}\,.
 \]
Symmetries related to an isotropic and homogeneous Universe impose that $q_\alpha =0=\pi_{\alpha\beta}$, and only the scalar dissipation $\Pi$ is possible \cite{marteensreview}.
The energy-momentum conservation law $T^{\alpha\beta}_{\,\,\,\,\,;\, \beta}=0$ reads
\begin{equation}\label{continuita1a}
     {\dot \rho}+\Theta (\rho+p+\Pi)=0\,,
\end{equation}
where the dot stands for the derivative with respect to the cosmic time and $\Theta = u^\alpha_{\,\,\,\, ; \alpha}$.

In an homogeneous and isotropic (spatially flat) Universe, the Einstein equations read \cite{zimdahl2000}
 \begin{equation}\label{EinsteinEq-FRW}
    H^2 = \frac{\kappa \rho}{3}\,, \quad\quad  {\dot H}=-\frac{\kappa}{2}(\rho +p+\Pi)\,,
 \end{equation}
where $H={\dot a}/{a}$ is the Hubble parameter and $\kappa = 8\pi G= 8\pi m_P^{-2}$ ($m_P \simeq 1.22 \times 10^{19}$GeV is the Planck mass).

The general equation for the evolution of the Hubble parameter is given by \cite{zimdahl2000}

\begin{widetext}

 \begin{equation*}
 \frac{\ddot H}{H}-\frac{{\dot H}^2}{\gamma H^2}\left(\frac{\rho+p}{T\partial \rho/\partial T}+1+\frac{\partial p}{\partial \rho}\right)
 -3{\dot H}\left[\frac{\rho+p}{T\partial \rho/\partial T}-1-\frac{\partial p}{\partial \rho}+\frac{1}{2}\left(\frac{\partial p}{\partial \rho}-c_s^2\right)\right]
 -\frac{9H^2\gamma}{2}\left[c_b^2+\frac{1}{2}\left(\frac{\rho+p}{T\partial \rho/\partial T}-1-\frac{\partial p}{\partial \rho}\right)-\right.
 \end{equation*}
 \begin{equation}\label{Hubble_evolution}
\left. -\frac{1}{2}\left(\frac{\partial p}{\partial \rho}-c_s^2\right)\right]
 -\frac{{(c^2_b)^.}}{2 c_b^2}\left(\frac{\dot H}{H}+\frac{3}{2}\gamma H\right)+\frac{1}{\tau}\left(\frac{\dot H}{H}+\frac{3}{2}\gamma H \right)=0\,,
\end{equation}
  \end{widetext}
where $\tau$ is the relaxation time (the mean free time of the relativistic particle) that in general is time dependent, $\tau=\tau(t)$,
 \[
 \gamma=1+\frac{p}{\rho}=1+\omega\,, \qquad
 c_s^2=\frac{\partial p}{\partial \rho}\Big|_{ad}=\frac{n}{\rho+p}\frac{\partial p}{\partial n}+\frac{T}{\rho+p}\frac{(\partial p/\partial T)^2}{\partial \rho/\partial T}\,,
 \]
$c_b^2\equiv \frac{\zeta}{(\rho+p)\tau}$ is the propagation velocity of viscous pulse \cite{hiscock}, and $\zeta$ is the coefficient of the bulk viscosity.
In a viscous medium the sound velocity is $v^2 = c_s^2+c_b^2 \leqslant 1$, i.e. the sound propagates with a subluminal velocity.
We also used $\Theta=3H$.

In the ultra-relativistic regime (radiation dominated era) one has
 \[
 p=\frac{\rho}{3}\,, \qquad \rho=\displaystyle{\frac{\pi^2 g_*}{30}T^4}\,,
 \]
where $g_*\simeq 106.7$ are the relativistic degrees of freedom, the field equation (\ref{Hubble_evolution}) reads \cite{zimdahl2000}
\begin{equation}\label{Hfieldradiation}
  \frac{\ddot H}{H}-2 \frac{{\dot H}^2}{H^2}-6 H^2 c_b^2+\frac{1}{\tau}\left(\frac{\dot H}{H}+2H\right)=0\,.
\end{equation}

\subsection{The universe with particle production}

In this Section we study the case of {\it isentropic} (or {\it adiabatic}) particle production, i.e. the number of particle is not conserved. Although the isentropic condition implies a constant entropy for particle, an entropy production is still present because of the enlargement of the phase space of the system due to the increasing of the number of perfect fluid particles.

If the number of particles is not conserved, one has to use \cite{zimdahl2000}
 \begin{equation}\label{dotN}
    \nabla_\mu N^\mu = {\dot n} + {\Theta} n = n\Gamma\,,
 \end{equation}
where $\Gamma={\dot N}/N$ and $N=na^3$ ($N$ is number of particles in the comoving volume $a^3$). $\Gamma > 0$ ($< 0$) means particles creation (annihilation). It is important to stress that
a nonvanishing particle production rate gives rise to an effective bulk pressure of the cosmic fluid.
Moreover, in a phenomenological description, $\Gamma$ is a input quantity whose expression is calculated from the
microphysics underlying the physical phenomena\footnote{A more realistic scenario requires to consider a Universe with two or more
fluids. For the case of two fluids, the continuity equations read
 \[
 {\dot \rho}_i+\Theta(\rho_i+p_i)=\sum_i \epsilon_i\Gamma_i \rho_i\,,
 \]
with $i=1, 2$ and $\epsilon_1= 1=-\epsilon_2$. Although these fluids do not separately satisfy the energy-momentum conservation, the total energy density does ${\dot \rho}+\Theta (\rho+p)=0$, with $\rho\simeq \rho_1+\rho_2$ and $p\simeq p_1+p_2$. In our case we are assuming for simplicity $\rho\simeq \rho_1 \gg \rho_2$.}. 

Using the Gibbs equation $Tds=d\displaystyle{\frac{\rho}{n}}-p\, d\displaystyle{\frac{1}{n}}$ and Eqs. (\ref{continuita1a})
and (\ref{dotN}) one gets
 \begin{equation}\label{dots}
    n T {\dot s}=-{\Theta} \Pi -(\rho+p)\Gamma\,.
 \end{equation}
The condition ${\dot s}=0$ (isentropic particle production) implies that viscous pressure is entirely determined by the particle production rate
 \begin{equation}\label{PiGamma}
    \Pi=-(\rho+p) \frac{\Gamma}{\Theta}\,.
 \end{equation}
It then follows \cite{zimdahl2000}
 \begin{equation}\label{raten-T}
    \frac{\dot n}{n}=-(\Theta-\Gamma)\,, \qquad \frac{\dot T}{T}=-(\Theta-\Gamma)\frac{\partial p}{\partial \rho}\,.
 \end{equation}
\begin{equation}\label{continutia2}
    {\dot \rho}=-(\Theta-\Gamma)(\rho+p)\,, \quad {\dot p}=-c_s^2(\Theta-\Gamma)(\rho+p)\,.
\end{equation}
In this description the cosmic substratum is a perfect fluid with varying particle number (and not a conventional dissipative fluid).

The combination of Eqs (\ref{EinsteinEq-FRW}) and (\ref{PiGamma}) yields
 \begin{equation}\label{Gamma/Theta}
    \frac{\Gamma}{\Theta}=1+\frac{2}{3\gamma}\frac{\dot H}{H^2}\,.
 \end{equation}
The time evolution of the Hubble expansion rate $H$, for $p=\rho/3$ is given by \cite{zimdahl2000}

 \begin{equation}\label{Htimeevolution1}
    \frac{\ddot H}{H}-\frac{{\dot H}^2}{\gamma H^2}\left(1+c_s^2+\frac{\partial p}{\partial \rho}\right)+3{\dot H} \left[1-\frac{1}{2}\left(\frac{\partial p}{\partial \rho}-c_s^2\right)\right]
    -\frac{9}{2}\gamma H^2\left(c_b^2 \frac{nsT}{\rho+p}-\frac{1}{2}\right) -
 \end{equation}
 \[
 -\frac{1}{2}\frac{(c_b^2)^.}{c_b^2}\left(\frac{\dot H}{H}+\frac{3}{2}\gamma H\right)+\frac{1}{\tau}\left(\frac{\dot H}{H}+\frac{3}{2}\gamma H\right)=0\,.
 \]
This is the counterpart of (\ref{Hfieldradiation}) for $\Gamma\neq 0$.
In the particular case $p=\rho/3$ and $ns=(\rho+p)/T$, Eq. (\ref{Htimeevolution1}) reduces to the following expression
 \begin{equation}\label{Htimeevolution1RD}
    \frac{\ddot H}{H}-\frac{5}{4}\frac{{\dot H}^2}{H^2}+3{\dot H} - 6H^2\left(c_b^2-\frac{1}{2}\right)+\frac{1}{\tau}\left(\frac{\dot H}{H}+2H\right)=0\,.
 \end{equation}

\section{Bulk Viscosity and Big Bang Nucleosynthesis Constraints}
\setcounter{equation}{0}


Since {\it pre} BBN epoch is a period of the Universe evolution not directly constrained by cosmological observations,
one has to require that the effects of bulk viscosity could manifest well before the BBN starts, hence for times $t_*$ (or temperatures $T_*$) lesser than those characterizing BBN: $t_* \ll t_{\text{BBN}}\sim (10^{-2}-10^{2})$sec (or  $T_* \gg T_{\text{BBN}}\sim (10-0.1)$MeV).
However, it is worth to investigate what constraints the BBN provides once bulk viscosity effects are taken into account in the Universe evolution
(this analysis is performed for a power law evolution of the scale factor $a(t) \propto t^{\varsigma}$).

During the BBN,  the relevant weak interactions are governed by the processes
 \[
 \nu_e+n \, \longleftrightarrow \, p+e^-\,, \quad
 e^++n \, \longleftrightarrow \, p+{\bar \nu}_e\,, \quad
 n \, \longleftrightarrow \, p+e^- +{\bar \nu}_e\,.
 \]
The weak interaction rate of particles in thermal equilibrium is given by \cite{bernstein,kolb}
 \begin{equation}\label{Lambdafin}
    \Lambda(T)\simeq q T^5+{\cal O}\left(\frac{\cal Q}{T}\right)\,,
 \end{equation}
where
 \[
 q=9.6 \times 10^{-46}\text{eV}^{-4}\,, \qquad {\cal Q}=m_n-m_p\ll T\,,
 \]
with $m_{n,p}$ are the neutron and proton masses.
Here
 \[
 \Lambda\equiv\Lambda_{\nu_e+n  \leftrightarrow  p+e^-}+\Lambda_{e^++n  \leftrightarrow  p + {\bar \nu}_e}+
 \Lambda_{n \leftrightarrow  p+e^- + {\bar \nu}_e}
 \]
is the sum of the weak interaction rates.

To estimate the primordial mass fraction of ${}^4 He$, one usually defines \cite{bernstein,kolb}
 \begin{equation}\label{Yp}
    Y_p\equiv \lambda \, \frac{2 x(t_f)}{1+x(t_f)}\,,
 \end{equation}
where $\lambda=e^{-(t_n-t_f)/\tau}$, $t_f$ and $t_n$ are the time of the freeze-out of the weak interactions and of the nuclesynthesis,
respectively, $\tau\simeq 887$sec is the neutron mean life, and $x(t_f)=e^{-{\cal Q}/T(t_f)}$ is the neutron to proton equilibrium ratio.
The function $\lambda(t)$ represents the fraction of neutrons that decay into protons in the time $t\in [t_f, t_n]$.
Deviations from $Y_p$ (generated by the variation of the freezing temperature $T_f$) are given by
 \begin{equation}\label{deltaYp}
    \delta Y_p=Y_p\left[\left(1-\frac{Y_p}{2\lambda}\right)\ln\left(\frac{2\lambda}{Y_p}-1\right)-\frac{2t_f}{\tau}\right]
    \frac{\delta T_f}{T_f}\,.
 \end{equation}
In the above equation we have set $\delta T(t_n)=0$ because $T_n$ is fixed by the deuterium binding energy \cite{torres,lambiaseBBN}.
By making use of the current estimation on $Y_p$ \cite{coc}
 \begin{equation}\label{Ypvalues}
 Y_p=0.2476\,, \qquad |\delta Y_p| < 10^{-4}\,,
 \end{equation}
one obtains
 \begin{equation}\label{deltaT/Tbound}
    \left|\frac{\delta T_f}{T_f}\right| < 4.7 \times 10^{-4}\,.
 \end{equation}
Exact solutions of Eq. (\ref{Hubble_evolution}) are, in general, extremely difficult to determine.
We make therefore the ansatz $\omega\approx p/\rho\approx \partial p/\partial\rho\approx 1/3$ (the Universe evolves isotropically) and that the scale
factor evolves, due to bulk viscosity, as $a(t) \propto t^{\varsigma}$, where $\varsigma=1/2+\delta$ with $\delta \ll 1$.
Therefore, the expansion rate of the Universe can be written in the form
 \begin{equation}\label{H-HGR}
    H=2\varsigma H_{GR}\,,
 \end{equation}
where $\quad H_{GR}=\frac{1}{2t}$ is the expansion rate obtained in General Relativity.
Imposing that the expansion rate of the Universe is equal of the interaction rate, $\Lambda\simeq  H$, one derives the freeze-out temperature
$T=T_f(1+\frac{\delta T_f}{T_f})$, with $T_f\sim 0.6$ MeV and
 \begin{equation}\label{deltaT/T}
    \frac{\delta T_f}{T_f} = \delta \frac{4\pi}{15}\sqrt{\frac{\pi g_*}{5}}\frac{1}{qm_P T_f^3}\simeq 1.0024 \delta\,.
 \end{equation}
Equations (\ref{deltaT/T}) and (\ref{deltaT/Tbound}) imply
 \begin{equation}\label{boubdc}
    \delta \lesssim \delta_{BBN}\,, \quad \delta_{BBN} \equiv 4.7 \times 10^{-4}\,, \quad \to \quad \varsigma\lesssim \frac{1}{2}+4.7 \times 10^{-4}
 \end{equation}
Hence, in the case $\Gamma=0$, from Eq. (\ref{Hfieldradiation}) it follows that  the relaxation time is given by
$\tau H = C \varsigma(1-2\varsigma)$, with $C=\left[\frac{17}{16}-\frac{3}{2}c_b^2\right]^{-1}\sim {\cal O}(1)$.
For $\Gamma\neq 0$, Eq. (\ref{Htimeevolution1RD}) implies $\tau H = \frac{1}{3}\left(1-\frac{1}{2\varsigma}\right)\left[c_b^2-\frac{1}{2}\left(1-\frac{1}{2\varsigma}\right)\right]^{-1}$ (this relation shows
$\tau H> 0$ provided $c_b> \frac{1}{2}\sqrt{\frac{2\varsigma-1}{\varsigma}}$ and $\varsigma>1/2$, i.e.
the internal fluid dynamics limits the amplitude of the effective viscous pressure \cite{zimdahl2000}).

The trace of the energy-momentum tensor (Eq. (\ref{energy-momentum-tensor})) reads
 \[
{\cal T}= T_\alpha^{\,\,\, \alpha}\simeq  3\Pi\,,
 \]
which implies ($R=-\kappa {\cal T}$)
 \[
 {\dot R}=-{\kappa} {\dot {\cal T}}=-3\kappa {\dot \Pi}\,.
 \]
Using the definition of the scalar curvature $R=-6({\ddot a}/a+{\dot a}^2/a^2)$ and $a\sim t^\varsigma$, one infers
 \begin{equation}\label{kPi-FRW}
 {\dot R}= \frac{12 \varsigma (2\varsigma-1)}{t^3}\,,
 \end{equation}
hence
 \begin{equation}\label{Pi(t)}
\Pi(t) = \frac{2\varsigma(2\varsigma-1)}{\kappa t^2}\,.
 \end{equation}
Both $\tau$ and $\Pi$ vanish in the limit $\varsigma=1/2$ ($\delta =0$).
Since
 \[
\Pi=2\varsigma(\varsigma-1/2)\rho \simeq \delta \rho \ll \rho\,,
 \]
the dissipative $\Pi$ term in (\ref{EinsteinEq-FRW}) represents indeed a tiny perturbation to energy density.

\section{Dark matter relic abundance}
\setcounter{equation}{0}

As we have seen in the previous Sections, scalar pressure can give rise to a different evolution of the early universe,
which may deviate from the  standard cosmology provided by general relativity. In this Section we wish to discuss
how these modifications of the standard cosmology affect the thermal relic abundance.

It is nowadays a very consolidate fact that our Universe is dominated
by dark matter (as well as by dark energy, responsible of the accelerated expansion of the Universe), whose ratio with the critical density satisfies the bounds \cite{spergel}
$0.092 \leq \Omega_{CDM} h^2 \leq 0.124$,
where $h=100$Km s$^{-1}$ Mpc$^{-1}$ is the Hubble constant.
Favorite candidates for non-baryonic cold dark matter seem to be the WIMPs (weakly interacting massive particles), which decoupled from the thermal plasma in the early Universe. The interest about these particles as dark matter follows from the fact that the abundance of WIMPs in chemical equilibrium in the early Universe agrees with the expected one in the context of cold dark matter.

According to standard cosmology and particle physics, the calculation of the relic densities of particles is based on the assumption that the
period of the Universe dominated by radiation began before the main production of relics and that the entropy of matter is conserved during this epoch and the successive one.
Once these assumptions are relaxed, a different relic density of particles is expected. In this scenario, therefore, any contribution to the energy density modifies the Hubble expansion rate, which reflects in a modification of the relic density values \cite{BD,fornengo}.

To account for the enhancement of the expansion rates, it is usual to write \cite{fornengo}
 \begin{equation}\label{H=AHGR}
    H(T)=A(T) H_{GR}(T)\,.
 \end{equation}
Here $H$ refers to the expansion rate of the cosmological model modified by bulk viscosity effects discussed in the previous Sections, while
$H_{GR}$ refers to expansion rate of standard cosmology.

The function $A(T)$ assumes values greater than 1 ($A(T) > 1$) at large temperatures, and $A(T)\to 1$ before BBN set up. The last is imposed by successful predictions of BBN on the abundance of primordial light elements.
The function $A(T)$ can be conveniently parameterized as \cite{fornengo} (see also \cite{gondolo})
\begin{equation}\label{A(T)T>Tre}
    A(T)=1+\eta\left(\frac{T}{T_f}\right)^\nu  \quad \stackrel{\eta\gg1}{\longrightarrow} \quad \eta\left(\frac{T}{T_f}\right)^\nu\,,
 \end{equation}
where $\eta$ and $\nu$ are free parameters that characterize the specific cosmological model, and $T_f\simeq 17.3$GeV is the normalization temperature: $A(T_f)=1+\eta$. The factor $1+\eta$ is hence the enhancement factor of the Hubble rate at the time of the WIMPs freeze-out\footnote{To be more precise, the function $A(T)$ is parameterized such that
for temperatures $T>T_{re}$ it assumes the form
\begin{equation}\label{A(T)}
    A(T)=1+\eta\left(\frac{T}{T_f}\right)^\nu \tanh \frac{T-T_{re}}{T_{re}}
 \end{equation}
while for $T\leq T_{re}$ it approaches to 1. The temperature $T_{re}$ denotes the temperature at which the Hubble rate reenters the standard rate of general relativity, while $T_f$ is a reference temperature, assumed in
\cite{fornengo} as the temperature at which the WIMPs dark matter freezes out in the standard cosmology ($T_f \simeq 17.3$GeV). In general, $T_f$ varies by varying the dark matter mass $m_\chi$. To avoid contradictions
with big bang nucleosynthesis, it is required $T_{re}\gtrsim 1$ MeV. Estimations carried out in \cite{fornengo} have been obtained by setting $T_{re}=1$ MeV. In the regime $T\gg T_{re}$, the function (\ref{A(T)}) behaves as (\ref{A(T)T>Tre}).}.


To explain the PAMELA data, together with $\Omega_\chi h^2 =\Omega h^2\big|^{\text{WMAP}}_{\text{CDM}}=0.1131\pm0.0034$ \cite{komatsu} for dark matter annihilation into $e^+ e^-$, the values of the parameter $\eta$ must be \cite{fornengo}
 \begin{equation}\label{etavaluesPAMELA}
    0 \lesssim \eta \lesssim 10^3\,.
 \end{equation}
In this range, the values of the WIMPs dark matter masses are
 \begin{equation}\label{massvaluesPAMELA}
10^2 \text{GeV}\lesssim m_\chi \lesssim 10^3\text{GeV}\,.
 \end{equation}
In particular, one has
 \begin{equation}\label{etavaluesPAMELAq}
 m_\chi\sim 10^2\text{GeV} \quad \mbox{for} \quad   0 \lesssim \eta \lesssim 1\,.
 \end{equation}
Let us now apply these results to an expanding Universe in presence of bulk viscosity, considering both the cases $\Gamma=0$ and $\Gamma\neq 0$.

\subsection{The case $\Gamma=0$}

Owing to the presence of a scalar pressure, a tiny deviation from standard cosmological evolution is expected.
Let us therefore assume that the Universe evolution during the radiation dominated era is governed by the scale factor $a\sim t^\varsigma$.  The Hubble expansion rate can be written as
 \begin{equation}\label{H=Hgr-ParticProduc1}
    H=2\varsigma H_{GR}\,, \quad H_{GR}=\frac{1}{2t}\,,
 \end{equation}
i.e. it corresponds to a overall boost of the Hubble expansion rate.
Eqs. (\ref{H=AHGR}) and (\ref{A(T)T>Tre}) imply $2\varsigma = 1+\eta$ and $\nu=0$.

In particular, if one takes into account BBN constraints, Eq. (\ref{boubdc}), then it follows
$\eta=2\delta \simeq 10^{-3}$, hence, according to (\ref{etavaluesPAMELAq}), the dark matter mass satisfying this condition is of the order $m_\chi\sim 10^2$GeV.

\subsection{The case $\Gamma\neq 0$}


The ansatz for the scale factor $a\sim t^\varsigma$
implies that during the radiation dominated era, the rates $H$ and $\Gamma$, related by Eq. (\ref{Gamma/Theta}), can be written in form
 \begin{equation}\label{Gamma3H}
    \Gamma=\left(1-\frac{1}{2\varsigma}\right) 3H\,.
 \end{equation}
Moreover, it also follows that
\begin{equation}\label{n-T-rho}
    n\propto a^{-3/(2\varsigma)}\,, \quad T\propto a^{-1/(2\varsigma)}\,, \quad \rho\propto a^{-2/\varsigma}\,.
\end{equation}
These relations imply $\rho\sim T^4$. 

The exponent $\varsigma$ in the scalar factor $a$ cannot be therefore arbitrary, but it is determined by
Eq. (\ref{Gamma3H}). For simplicity we consider the case $\Gamma = \alpha H$, with $\alpha\neq 3$. One gets $\varsigma=\frac{3}{2(3-\alpha)}$.
For $0 \leqslant \alpha < 3$, the Universe hence evolves more rapidly ($a\sim t^{\frac{3}{2(3-\alpha)}}$) with respect to the standard case ($a\sim t^{1/2}$), leading to an enhancement of the Hubble expansion rate. The latter is related to the standard one, $H_{GR}$, by the relation
(\ref{H=Hgr-ParticProduc1}). Comparing with (\ref{H=AHGR}) and (\ref{A(T)T>Tre}), one hence obtains
 \begin{equation}\label{q=1/2}
   \eta=\frac{\alpha}{3-\alpha}\,, \quad \nu=0\,.
 \end{equation}

According to (\ref{massvaluesPAMELA}), the dark matter mass is the order of $m_\chi \sim (10^2-10^3)$GeV for $\alpha\approx 3$, i.e. $\eta \gg 1$,
while $\alpha \ll 1$, i.e. $\eta <1$, yields $m_\chi \sim 10^2$GeV, as follows from (\ref{etavaluesPAMELAq}).

\subsection{A general solution for $\Gamma=0$ and $\Gamma\neq 0$}

In this Section we discuss a more general solution to Eq. (\ref{Hfieldradiation}). During the pre-BBN, the expansion rate of the Universe can be parametrized as
 \begin{equation}\label{Hgeneral}
   H(t) = H_L \left(\frac{t}{t_L}\right)^\Upsilon\,,
 \end{equation}
where $H_L$, $t_L$ and $\Upsilon$ are constants. The relation between the cosmic time and the temperature $T$ follows from the field equation (\ref{EinsteinEq-FRW}). One infers
 \begin{equation}\label{t-Trelation}
   t=t_L \left(\frac{8 \pi^3 g_*}{90}\right)^{1/2\Upsilon}\left(\frac{T^2}{m_P H_L}\right)^{1/\Upsilon}\,.
 \end{equation}
It is straightforward to determine the enhancement factor $A(T)$. Writing $H = A(T) H_{GR} = A(T)/ (2t) $, one immediately obtains
 \begin{equation}\label{Ageneral}
   A(T)=\eta \left(\frac{T}{T_f}\right)^\nu\,, \qquad \eta\equiv 2H_L t_L \left[\left(\frac{4\pi^3 g_*}{45}\right)^{1/2}\frac{T_f^2}{m_P H_L}\right]^{(\Upsilon+1)/\Upsilon}\,, \quad \nu \equiv \frac{2(\Upsilon+1)}{\Upsilon}
 \end{equation}
By fixing parameters entering into (\ref{Ageneral}) one can obtain the conditions for which
Eqs. (\ref{etavaluesPAMELA}), (\ref{massvaluesPAMELA}) and (\ref{etavaluesPAMELAq}) are fulfilled.
Therefore, dark matter particles with mass of the order of $(10^2-10^3)$GeV can be accommodate in this model.

The characteristic time $\tau(t)$ can be obtained by substituting (\ref{Hgeneral}) into Eq. (\ref{Hfieldradiation})
 \[
  \tau H = \frac{1+\frac{\Upsilon}{2H_Lt_L}\left(\frac{H_L}{H}\right)^{1+1/\Upsilon}}{3c_b^2+\frac{\Upsilon(\Upsilon+1)}{2t_L^2 H_L^2}\left(\frac{H_L}{H}\right)^{2+1/\Upsilon}}\,.
  \]
{\it Mutatis mutandis}, similar results follow for $\Gamma\neq 0$ by using (\ref{Htimeevolution1RD}), i.e. in the case of particles production.

\section{Gravitational Baryogenesis}
\setcounter{equation}{0}

Let us discuss the effects of the bulk viscosity on the generation of baryon asymmetry.
The origin of the baryon number asymmetry in the Universe is still an unsolved problem \cite{kolb,reviewBaryo}. The success of the big bang nucleosynthesis
\cite{Copi,burles} and the observations of Cosmic Microwave Background anisotropies (combined with the large structure of the
Universe \cite{wmap,bennet}) show that the parameter characterizing such an asymmetry  is of the order
 \begin{equation}\label{etaexp}
\eta_B \equiv \frac{n_B-n_{\bar B}}{s}\lesssim (9.2 \pm 0.5)\,\, 10^{-11}\,,
 \end{equation}
where $n_B$ ($n_{\bar B}$) is the baryon (antibaryon) number
density, and $s$ the entropy of the Universe ($s=2\pi^2g_{*}T^3/45$).

In this paper we refer to the model proposed in \cite{steinhardt}. Here the baryon number asymmetry
is due to a dynamical breaking of CPT (and CP) symmetry that is generated by the coupling of
the derivative of the Ricci scalar curvature
$R$ with the baryon current $J^\mu$ \cite{steinhardt}
\begin{equation}\label{riccicoupling}
  \frac{1}{M_*^2}\int d^4x \sqrt{-g}J^\mu\partial_\mu R\,,
\end{equation}
where $M_*$ is the cutoff scale characterizing the effective
theory. If there exist interactions that violate the baryon number
$B$ in thermal equilibrium, then a net baryon asymmetry can be
generated and gets frozen-in below the decoupling temperature
$T_D$. Here the thermal equilibrium of interaction that violate $B$ does not refer to those interactions
that generate the bulk viscosity, as discussed in the Introduction.
From (\ref{riccicoupling}) it follows
 \[
M_*^{-2} (\partial_\mu R)J^\mu=M_*^{-2} {\dot R}(n_B-n_{\bar B})\,,
 \]
where ${\dot R}=dR/dt$. The effective chemical potential for
baryons and antibaryons is $\mu_B={\dot R}/M_*^2 =-\mu_{\bar B}$,
and the net baryon number density at the equilibrium turns out to
be (as $T\gg m_B$, where $m_B$ is the baryon mass) $n_B=g_b\mu_B
T^2/6$. $g_b\sim {\cal O}(1)$ is the number of intrinsic degrees
of freedom of baryons. The baryon number to entropy ratio, that defines
the baryon asymmetry, is therefore \cite{steinhardt}
\begin{equation}\label{nB/s}
 \eta_B= \frac{n_B}{s}\simeq -\frac{15g_b}{4\pi^2g_*}\frac{\dot R}{M_*^2 T}\Big|_{T_D}\,.
\end{equation}
where $g_*\sim 106.7$. The trace of field equations gives 
 \[
 {\dot R}=-{\kappa} {\dot {\cal T}}=-{\kappa}^{3/2}(3\omega-1)(\omega+1)\rho^{3/2}-3\kappa {\dot \Pi}\,.
 \]
Deviations from the standard cosmology provide a non vanishing ${\dot R}$, so that a net baryon
asymmetry can be be generated also during the radiation dominated era ($\omega=1/3$).
To compute ${\dot \Pi}$ we confine ourselves to the case\footnote{
In the case $\Gamma=0$ the expression of $\Pi$ is \cite{zimdahl2000} $\kappa \Pi = -3\gamma H^2 -2 {\dot H}$, and its time derivative turns out to be
 \begin{equation}\label{Piderivative2}
   \kappa {\dot \Pi} = -2 {\ddot H}-6 H {\dot H}\left(1+\frac{\partial p}{\partial \rho}\right)+9\gamma H^3\left(c_s^2-\frac{\partial p}{\partial \rho}\right)\,.
 \end{equation}
Assuming that during the pre-BBN the Universe evolves according to (\ref{Hgeneral}), direct calculations show that
a net baryon asymmetry can be generated, and that the observed value (\ref{etaexp}) can be obtained for appropriate fine tuning
of free parameters $\{\Upsilon, H_L, t_L\}$. Instead, in the case in which the scale factor evolves as $\sim t^\varsigma$, and BBN constraints are taking into account, a net baryon asymmetry is still generated, but it is too small with respect to the observed value (\ref{etaexp}).}  $\Gamma\neq 0$.

Using Eqs (\ref{PiGamma}) and (\ref{continutia2}), and taking $\Gamma\sim H$ and $M_* = m_P/\sqrt{8\pi}$, it follows that
 \[
 {\dot \Pi}=\frac{32}{27} \sqrt{8\pi} \,\,\frac{\rho^{3/2}}{m_P}\,,
 \]
so that the baryon number asymmetry (\ref{nB/s}) is
 \begin{equation}\label{etapartproduction}
    \eta_B=\frac{40(8\pi)^{5/2}}{3}\frac{g_b \pi^2 g_*^{1/2}}{(30)^{3/2}}\frac{T_D^5}{M_*^2 m_P^3}\simeq 5.24 \times 10^{4}\left(\frac{T_D}{m_P}\right)^5\,
 \end{equation}
The net baryon asymmetry (\ref{etaexp}) follows for decoupling temperature of the order $T_D\sim 10^{-3}m_P$, i.e. at GUT scales.

\section{Conclusion}

In this paper we have analyzed some cosmological consequences of imperfect fluids, which are characterized
by an energy-momentum tensor that contains, in general, the bulk viscosity pressure, the energy flux and the anisotropic stress.
The bulk viscosity term, which we were mainly interested in, arises in mixture of matter (either of different species as in a radiative fluid or
of the same species but with different energies as in a Maxwell-Boltzmann gas) or particle production.

Deviations from the standard cosmology, induced by scalar pressure, have important impact on the problem of dark matter in the Universe.
The recent results of PAMELA experiment may be interpreted as identifying relic particles as the main constituents of dark matter.
The scalar pressure generates in the early Universe an enhancement of the Hubble expansion rate,
giving rise to thermal relics with a larger abundance. We have estimated that the mass of the WIMPs dark matter,
required to explain the excess of positron events found in PAMELA experiment, must be of the order of $10^2$GeV.

\vspace{0.2in}

A.I. acknowledges the Czech Science Foundation (GA\v{C}R), Contract No. 14-07983S, for partial support. G.L. thanks the ASI (Agenzia Spaziale Italiana) for partial support through the contract ASI number I/034/12/0.

\end{document}